# Bandwidth Enhancement in Multimode Polymer Waveguides Using Waveguide Layout for Optical Printed Circuit Boards

**Jian Chen, Nikos Bamiedakis, Peter Vasil'ev, Richard V. Penty, and Ian H. White**
*Electrical Engineering Division, Department of Engineering, University of Cambridge, 9 JJ Thomson Avenue, Cambridge, CB3 0FA, UK*
*Author e-mail address: jc791@cam.ac.uk*

**Abstract:** Dispersion studies demonstrate that waveguide layout can be used to enhance the bandwidth performance of multimode polymer waveguides for use in board-level optical interconnects, providing >40 GHz×m without the need for any launch conditioning.

## 1. Introduction

Short-reach optical interconnects have attracted significant interest for use in datacommunication links within data centre and supercomputer environments [1]. Due to the ever increasing demand for higher interconnection bandwidth, conventional copper-based electrical interconnects impose a performance bottleneck owing to their inherent disadvantages. Optical technologies provide a promising solution to this issue owing to their advantages over their electrical counterparts: higher bandwidth, lower crosstalk and improved power efficiency [2]. Multimode polymer waveguides are considered to be an excellent candidate for use in board-level optical interconnects enabling the formation of electro-optic (EO) printed circuit boards (PCBs). Siloxane polymer materials in particular, enable the direct integration of such waveguides onto standard PCBs owing to their good thermal and mechanical properties [3]; while the large core dimensions employed (typically 30 to 70 μm) offer relaxed alignment tolerances, thus enabling low-cost system assembly with pick-and-place tools [4].

However, the highly-multimoded nature of these waveguides has raised important questions concerning their bandwidth performance and their ability to support very high on-board data rates (>40 Gb/s). We have recently presented bandwidth studies on a 1 m long polymer multimode spiral waveguide using time domain measurements under various launch conditions [5], and demonstrated the potential of transmitting very high data rates (>100 Gb/s) over a single waveguide channel using refractive index engineering and launch conditioning [6]. In this paper, we present a novel approach to achieve bandwidth enhancement in such waveguides by using waveguide layout and exploiting the mode-filtering properties of passive waveguide components such as waveguide bends and crossings. Mode-filtering schemes have been used in multimode fibre systems [7] and mode-selective ring resonators [8] and couplers [9] for few-moded systems to increase data capacity. A similar approach is proposed here in the context of highly-multimoded board-level optical interconnects using simple passive waveguide components such as bends and crossings, providing a useful tool in the design of optical backplanes. Dispersion studies carried out on the waveguide bends and crossings show that the suppression of higher-order modes in these waveguides components results in a bandwidth enhancement up to ~1.5 times over similar straight waveguides and indicate that intelligent waveguide layout, together with the appropriate choice of refractive index profile, can ensure bandwidth-length product (BLP) >40 GHz×m to support very high on-board data rates while maintaining low loss performance and without the need for any launch conditioning at the waveguide input. The results highlight the potential of this technology for use in high-performance board-level optical interconnects.

## 2. Waveguide samples and experimental setup

Two waveguide samples with different refractive index profiles are fabricated from siloxane materials (core: Dow Corning® WG-1020 Optical Waveguide Core; cladding: XX-1023 Optical Waveguide Clad) on 8-inch silicon substrates using conventional photolithography. Various waveguide components are fabricated on each waveguide sample, including waveguide bends and crossings. The waveguide bends comprise four 90° bends, two of them with a constant radius (17 mm), and the other two with a varying radius of curvature from 5 to 20 mm [Fig 1(a)]. The waveguide crossings comprise of 4 waveguide bends and a variable number of 90° crossings from 1 to 80 [Fig. 1(b)]. Fig. 1(c) and (d) show images of the two waveguide components illuminated with red light. The refractive index (RI) profile of the two waveguide samples (denoted WG A and WG B) are measured using the refractive near field method and are illustrated in Fig. 2(a), while Fig. 2(b) summaries their characteristics. The two different profiles are generated by adjusting the fabrication parameters and waveguide thickness [10].

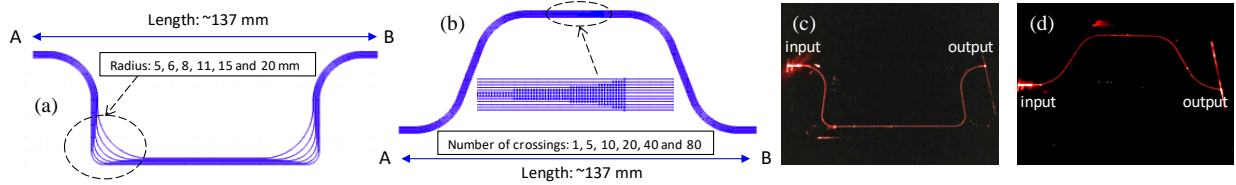

Fig. 1 Schematic of the waveguide components (a) bends, (b) crossings and top view of the waveguide (c) bend (R = 8 mm), (d) crossing (single crossing) illuminated with red light.

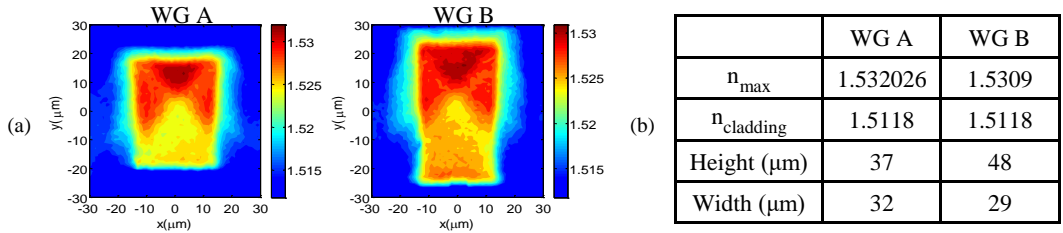

Fig. 2 (a) Measured RI profiles of the two waveguide samples at 678 nm and (b) summary of their characteristics.

|  | WG A | WG B |
| --- | --- | --- |
| $n_{max}$ | 1.532026 | 1.5309 |
| $n_{cladding}$ | 1.5118 | 1.5118 |
| Height (μm) | 37 | 48 |
| Width (μm) | 32 | 29 |

Dispersion measurements are carried on these waveguide components using a short pulse laser source and an autocorrelator. The experimental setups used are shown in Fig. 3(a) and 3(b). A femtosecond erbium-doped fibre laser operating at ~1574 nm (TOPTICA FFS) is utilised as the light source and a frequency-doubling crystal (MSHG1550-0.5-1) is employed to generate second harmonic pulses at wavelength of ~787 nm. The output light is coupled into a cleaved short 50/125 μm MMF patchcord via a 10× microscope objective (NA = 0.25) while the other end of the 50/125 um MMF patchcord is cleaved and mounted on a translation stage to enable butt-coupling into the waveguides. A near field image and the far field intensity of the 50/125 μm MMF input are shown in Fig. 3(c) and 3(d) respectively. The launch employed in the measurements is neither highly restricted or overfilled corresponding therefore, to a relatively realistic launch condition that could be encountered in a real-world system. The output light from the waveguide is collected with a 16× microscope objective (NA = 0.32) and is coupled to an autocorrelator to detect the transmitted optical pulses. The width of the received signal pulses are estimated from the autocorrelation traces using curve fitting with common pulse shapes (i.e. sech$^2$ or Lorentzian). The frequency response of the waveguide components is extracted by taking the Fourier transform of the back-to-back and waveguide links and subtracting one from the other. As a result, the -3 dB bandwidth of the waveguide components can be determined.

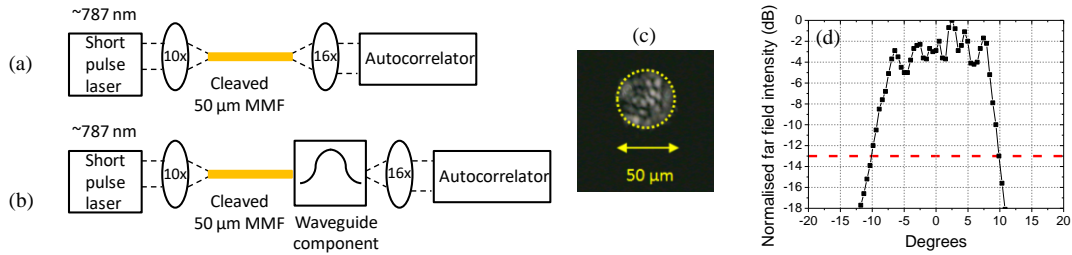

Fig. 3 Experimental setup used in the dispersion measurements for (a) the back-to-back and, (b) the waveguide link, (c) near field image and (d) far-field intensity at the output of the 50/125 μm MMF input fibre (5% intensity value noted in red dashed line).

A similar setup is used for the measurement of the loss performance of the components. The loss measurements are conducted at 850 nm using a multimode VCSEL source. A pair of microscope objectives is employed to couple the light into the short 50/125 um MMF patchcord. The cleaved end of the input fibre is butt-coupled to the waveguide input facet while at the waveguide output, a 16× microscope objective is used again to collect the output light and focus it onto an optical power meter head (HP 81525A). For each measurement, the position of the input fibre is adjusted with a precision translation stage to maximise the power transmission through the waveguide component under test. Similar loss and dispersion measurements are carried out on two straight waveguide samples (9.1 cm and 7.4 cm in length) with similar refractive index profiles and dimensions to allow direct comparison of the bandwidth and loss performance of the waveguide components under study.

## 3. Experimental results and discussion

Fig. 4(a) shows the estimated BLP for the waveguide bends as a function of the bend radius for the two waveguide samples with the different RI profile under the 50/125 μm MMF launch. The BLP values obtained for the respective straight waveguides under similar launch conditions are also shown in the plots for comparison: WG A~39 GHz×m, WG B~40 GHz×m. As expected, the waveguide bandwidth increases for decreasing bend radius due to the increased suppression of the higher-order modes along the bends. The insertion loss measurement shown in Fig. 4(b) confirms this observation. The insertion loss of two straight waveguides with length from point A to B (Fig. 1) is estimated to be 1.2 dB and 1.1 dB respectively under a similar type of launch (50/125 μm MMF) and is also noted in the plots with straight lines. As a result, it is found that a 1.55 times bandwidth improvement (BLP > 60 GHz×m) over the straight waveguides can be achieved for a bend radius of 5 mm, with a loss penalty of ~1.9 dB. A smaller loss penalty is introduced for slightly larger bends, for example, an 11 mm bend radius ensures a BLP > 50 GHz×m with an additional loss of ~0.7 dB.

A similar study is carried on the waveguide crossings. Fig. 4(c) shows their estimated BLP as a function of the number of crossings for the 50/125 μm MMF launch. The bandwidth increases for a larger number of crossings but its value saturates for more than 10 crossings. This behaviour matches the non-linear increase in loss performance in waveguide crossings reported in [11] and observed here as well [Fig. 4(d)], as higher-order modes are primarily attenuated in the first initial crossings. The results indicate that a bandwidth enhancement up to 1.25 times can be achieved therefore with a relatively low number of crossings ~10, which corresponds to an additional loss of ~2 and 1.6 dB for the samples WG A and WG B respectively. It should be noted that the difference in bandwidth and loss performance observed between the straight waveguides and the waveguide with a single crossing is due to the presence of the input/output waveguide bends in the component under study [Fig. 1(b)].

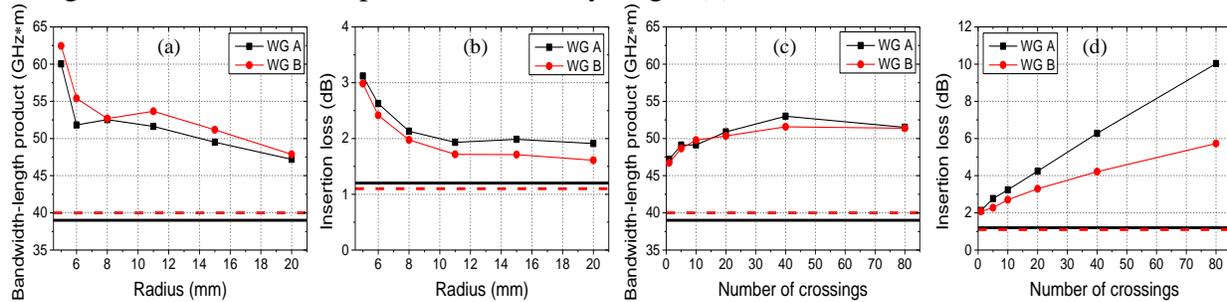

Fig. 4 (a) BLP of 90° bends versus bend radius, (b) insertion loss of the 90° bends versus bending radius, (c) BLP of 90° crossings versus number of crossing and (d) insertion loss of the 90° crossing versus number of crossing for a 50/125 μm MMF (BLPs and losses of respective straight waveguides noted: WG A in black solid line, WG B in red dashed line).

## 4. Conclusions

Dispersion measurements are conducted on polymer multimode waveguide bends and crossings with different refractive index profiles. The results show the potential of increasing the bandwidth performance of these multimode waveguides by ~1.5 times employing bend structures and waveguide crossings. Sufficient bandwidth for supporting >40 Gb/s on-board data transmission can be achieved while maintaining low loss transmission characteristics. The results demonstrate the importance of intelligent layout in optical backplanes based on this multimode polymer waveguide technology and highlight its potential in board-level optical interconnects.

## 5. Acknowledgements

The authors would like to acknowledge Dow Corning for providing the waveguide samples and EPSRC for supporting the work. Additional data related to this publication is available at the University of Cambridge data repository (https://www.repository.cam.ac.uk/handle/1810/251390).